\documentclass[journal=jacsat,manuscript=article]{achemso}

\usepackage[version=3]{mhchem} 
\usepackage{paralist}
\usepackage{booktabs,siunitx}
\usepackage{array}
\usepackage{dcolumn}
\newcommand{\mc}{\multicolumn}
\newcolumntype{H}{>{\setbox0=\hbox\bgroup}c<{\egroup}@{}}
\usepackage{amssymb}

\author{Juan Camilo Zapata}
\author{Laura K. McKemmish}
\email{l.mckemmish@unsw.edu.au}
\affiliation[University of New South Wales]
{School of Chemistry, University of New South Wales, 2052, Sydney}

\title[An \textsf{achemso} demo]
  {On the Computation of Dipole Moments: A Recommendation on the Choice of the Basis Set and the Level of Theory}

\usepackage[utf8]{inputenc}
\DeclareUnicodeCharacter{2212}{-}
\begin{document}







\begin{abstract}
    Together with experimental data, theoretically predicted dipole moments represent a valuable tool for different branches in the chemical and physical sciences. With the diversity of levels of theory and basis sets available, a reliable combination must be carefully chosen in order to achieve accurate predictions. In a recent publication (J. Chem. Theory Comput. 2018, 14, 4, 1969–1981), Hait and Head-Gordon  took a first step in this regard by providing recommendations on the best density functionals suitable for these purposes. However, no extensive study has been performed to provide recommendations on the basis set choice. Here, we shed some light into this matter by evaluating the performance of 38 general-purpose basis sets of single up to triple zeta-quality, when coupled with nine different levels of theory, in the computation of dipole moments. The calculations were performed on a data set with 114 small molecules containing second- and third-row elements. We based our analysis in regularised root mean square errors (regularised RMSE), where the difference between the calculated $\mu_{calc}$ and benchmark $\mu_{bmk}$ dipole moment values is derived as $(\mu_{calc}[D] - \mu_{bmk}[D])/(max(\mu_{bmk}[D],1[D]))$. This procedure ensures relative errors for ionic species and absolute errors for species with small dipole moment values. Our results indicate that the best compromise between accuracy and computational efficiency is achieved by performing the computations with an augmented double zeta-quality basis set (i.e. aug-pc-1, aug-pcseg-1, aug-cc-pVDZ) together with a hybrid functional (e.g. $\omega$B97X-V, SOGGA11-X). Augmented triple-zeta basis sets could enhance the accuracy of the computations, but the computational cost of introducing such a basis set is substantial compared with the small improvement provided. These findings also highlight the crucial role that augmentation of the basis set with diffuse functions on both hydrogen and non-hydrogen atoms plays in the computation of dipole moments.
\end{abstract}

\section{Introduction}
Molecular dipole moments play a crucial role for many chemical properties. A single dipole moment arises from the difference in electronegativity between the two atoms that constitute a chemical bond. The three-dimensional nature of molecules allows individual dipole moments to act along different directions, which results in the addition or cancellation of these quantities depending upon the molecular geometry \cite{36Si,65Da,07Pi}. Therefore, molecular dipole moments can deliver insight into the study of molecular structures. But it is not only in the elucidation of molecular structures that dipole moments play a crucial role. Strong dipole-to-dipole interactions are translated into higher melting and boiling points for a given compound. This is due to the strong intermolecular interactions that make it harder to separate the molecules in the compound \cite{13MoDa}. Dipole moments also play a crucial role in vibrational and rotational spectroscopy. A molecule with a permanent dipole moment will absorbs radiation from the microwave region of the spectrum leading to rotational transitions, whereas the arising of a dynamic dipole moment by infrared radiation, derives in vibrational transitions for a given molecule \cite{16Si, 13EnRe_Chap18, 13EnRe_Chap19, 97McSi_Chap5, 97McSi_Chap13}. 

With the diverse applications of dipole moments, the need of high-quality measurements of these quantities is imperative as it can serve as reference data for further experiments. State-of-the-art quantum chemistry packages can provide a reliable way to generate these results, as they are capable of calculating dipole moments following a straightforward black-box procedure. However, considering the volume of basis sets and levels of theory available, the choice of a particular model chemistry -- i.e. a level of theory and basis set combination -- can be overwhelming. 

A first step in addressing this issue was recently made by Hait and Head-Gordon \cite{18HaHe}, who calculated dipole moments for a data set containing 152 small molecules at the CCSD(T) level of theory, extrapolated to the complete basis set (CBS) limit. These benchmarking results were used to test the performance of 88 popular density functionals from all levels of Jacob's Ladder. They found remarkable performance delivered by hybrid and double-hybrid functionals, whereas local functionals, in general, were found to deliver lower performance. 

As dipole moments highly depend upon the tail region of the wave function, the basis set used in the computations must provide an accurate description of this region \cite{17NaJe, 99HaKlHe}. This requirement can be fulfilled by augmenting the basis set with diffuse functions, which are normally $s$- or $p$-functions with very small exponents that force them to decay slowly with the distance from the nucleus \cite{13Je_Book}. This feature allows the basis set to cover a wider range and therefore better describes the region far from the nucleus. An example where this improvement can be observed, is the study conducted by Halkier and co-workers \cite{99HaKlHe} whom demonstrated that the inclusion of diffuse functions in the basis set, results in accurate calculated dipole moments. Furthermore, they showed that diffuse functions provide a faster basis set convergence of the SCF part of the dipole moment in the calculations. More recently, Hickey and co-workers \cite{14HiRo} tested the influence of the Dunning's cc-pV\{D,T\}Z and aug-cc-pV\{D,T\}Z bases and the Sadlej pVTZ basis set in the calculation of dipole moments. They found that basis sets of triple zeta-quality augmented with diffuse functions (i.e. aug-cc-pVTZ) deliver the most accurate results, whereas unaugmented basis sets resulted in considerable lower performance. Conway and co-workers \cite{20CoGoPo} used the Dunning's aug-cc-pCV$n$Z basis set family to propose a strategy to calculate highly accurate dipole moments. Using a two-point formula, they were able to produce high-quality dipole moments by extrapolating the energies calculated with small basis sets, at a substantial reduction of computational effort. 

Though the knowledge of augmenting the basis set with diffuse functions in dipole moment calculations appears to be well established in the community, a comprehensive study to test the performance of different basis set families in these calculations has not been performed. The fact that nearly all general-purpose basis sets available possess an augmented version with diffuse functions present, implies that the basis set family quality will also have an influence in the calculations. Here, we aim to fill this gap by testing the performance of 38 general-purpose basis sets of single up to triple zeta-quality coupled with nine different levels of theory. The calculations are performed across a data set containing 114 molecules from second- and third-row elements. 

This paper is organised as follows: Section \ref{method} describes the method used for our analysis. This includes an explanation of the data set, as well as the analysis procedure and computational details followed throughout the study. Section \ref{results} provides the results with a discussion made in terms of basis set zeta-quality. This section finishes with our recommendations on the best basis set and level of theory to be used in the computation of dipole moments. Finally, a summary of our findings and some concluding remarks are presented in Section \ref{conclusion}.

\section{Method}
\label{method}

\subsection{Data Set}

We considered 114 molecules containing second- and third-row elements. These molecules were taken from the 152 small molecules studied by Hait and Head-Gordon \cite{18HaHe}, including their original geometries. Molecules containing Li, Be, B, Na and Al atoms were removed as some basis sets (pVTZ and ma-$n$) tested in this study were not available for those elements. 

We assessed the performance of 38 general-purpose basis sets, from single up to triple zeta-quality. The choice of these basis sets was representative in order to cover a wide range of common general-purpose basis sets. The basis sets considered were: \begin{compactenum}
    \item the Slater-mimicking minimal basis set STO-6G \cite{69HeStPo,70HeDiSt};
    \item the Pople-style basis sets 6-31G \cite{71DiHePo,72HeDiPo,75DiPo,77BiPo,82FrPiHe} and 6-311G \cite{80KrBiSe,80McCh}. The bases 6-31+G \cite{83ClChSp, 92GiJoPo, 87SpClRa} and 6-311+G \cite{83ClChSp, 92GiJoPo, 87SpClRa} augmented with diffuse functions on non-hydrogen atoms only. The bases 6-31G* \cite{73HaPo, 82FrPiHe} and 6-311G* \cite{80KrBiSe, 82FrPiHe} augmented with polarisation functions on non-hydrogen atoms only. The bases 6-31+G* \cite{71DiHePo, 72HeDiPo, 73HaPo, 75DiPo, 77BiPo, 82FrPiHe, 83ClChSp, 87SpClRa, 92GiJoPo} and 6-311+G* \cite{80KrBiSe, 80McCh, 82FrPiHe, 83ClChSp, 87SpClRa, 92GiJoPo} augmented with both diffuse and polarisation functions on non-hydrogen atoms, as well as the bases 6-31++G** \cite{71DiHePo, 72HeDiPo, 73HaPo, 75DiPo, 77BiPo, 82FrPiHe, 83ClChSp, 87SpClRa}, 6-311++G** \cite{80KrBiSe, 80McCh, 82FrPiHe, 83ClChSp, 87SpClRa} and 6-311++G(2df,2pd) which include diffuse and polarisation functions in both hydrogen and non-hydrogen atoms;
    \item the Jensen polarisation consistent basis sets  (pc-$n$ \cite{01Je,02Je,04JeHe,07Je} and aug-pc-$n$ \cite{01Je,02Je,02Je_b,04JeHe,07Je}, $n$ = 0, 1 and 2) and their segmented version (pcseg-$n$ \cite{14Je} and aug-pcseg-$n$ \cite{14Je}, $n$ = 0, 1 and 2);
    \item the Dunning correlation consistent basis sets (cc-pV$n$Z \cite{89Du,93WoDu,94WoDu,11PrWoPe} and aug-cc-pV$n$Z \cite{89Du,92KeDuHa,93WoDu,94WoDu,11PrWoPe} $n$ = D, T);
    \item the Ahlrichs-Karlsruhe basis set family def2-$n$ \cite{05WeAh}, the ma-$n$ \cite{05WeAh,11ZeXuTr} bases augmented with diffuse functions on non-hydrogen atoms and the def2-$n$D bases augmented with diffuse functions on both hydrogen and non-hydrogen atoms \cite{05WeAh, 10RaFu} ( $n$ = SVP, TZVP and TZVPP); and
    \item the Sadlej pVTZ basis set \cite{88Sa,91Sa,91SaUr}, a triple-zeta basis set developed specifically to calculate dipole moments. 
\end{compactenum}

To test the interaction of the basis set and the level of theory, the calculations were performed with nine different levels of theory. We included the benchmark wave function methods Hartree-Fock (HF) \cite{30Fo} and MP2 \cite{34MoPl}. The density functional methods chosen were based on recommendations of Hait and Head-Gordon \cite{18HaHe} namely: M06 \cite{08ZhTr}, SOGGA11-X \cite{11PeTr}, $\omega$B97X-V \cite{14MaHe}, B2PLYP \cite{09ZhXuGo} and DSD-PBEPBE-D3 \cite{13KoMa}. We also included B3LYP \cite{93Be,94StDeCh} for comparison due to its popularity, especially for infrared spectroscopy, and included the dispersion corrected  B3LYP-D3 \cite{10GrAnEh} (not considered by Hait and Head-Gordon) since dispersion is known to be important in accurate dipole moment predictions. 

\subsection{Analysis Approach}

Our paper aims to identify suitable modest basis set and level of theory combinations, that provide high quality dipole moments for a given reasonable calculation time.

\begin{figure}[ht]
    \centering
    \includegraphics[width=0.48\textwidth]{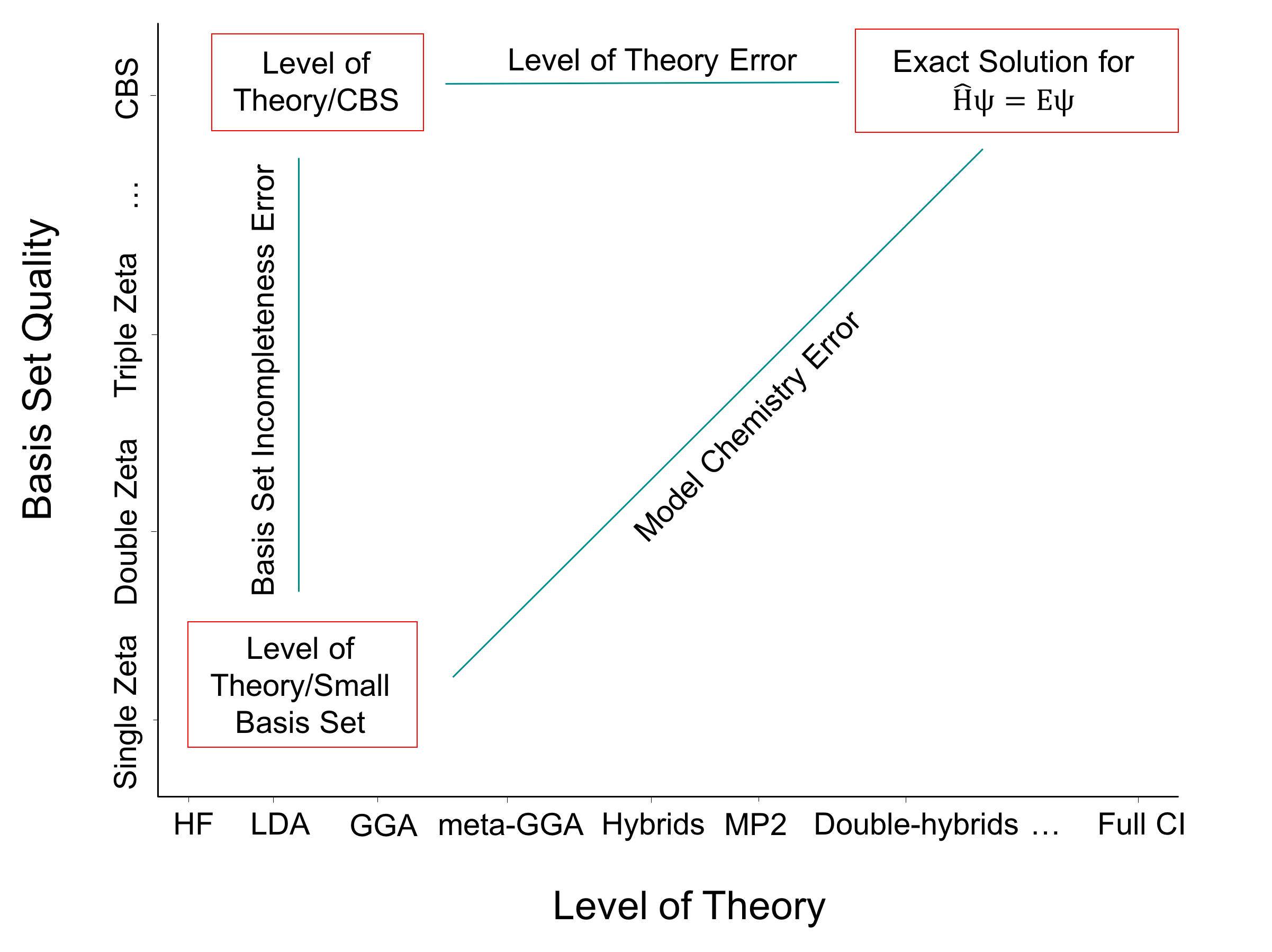}
    \caption{Combined Jacob's Ladder and Pople diagram for \textit{ab initio} quantum chemistry calculations.}
    \label{comp-errors}
\end{figure}

Figure \ref{comp-errors} presents our main tool to assess the accuracy of our calculations. The $y$ axis is basis set size while the $x$ axis uses a generalised `Level of Theory' ordering both wave function methods and DFT functionals in order of computational time, thereby merging the traditional Pople diagram \cite{65Po} with Jacob's Ladder \cite{19GoMe}. This figure clearly labels with lines the three important properties for determining the accuracy of our computational quantum chemistry calculations - basis set incompleteness error (BSIE), level of theory error (LoTE) and model chemistry error (ModChem error). Each of the three metrics can be calculated from the difference between two of the three calculations shown in red boxes in Figure \ref{comp-errors}:  a modest calculation (i.e. moderate basis set and level of theory) whose accuracy is to be determined, a complete basis set calculation with the same level of theory and the exact solution to the Schrodinger equation.   

For practical calculations, basis set and level of theory combinations should be chosen such that the basis set incompleteness error and level of theory error are comparable, with the least time consuming contribution having the lower error; e.g. if choice of level of theory is more important for timings than basis set choice, then the basis set should be increased in size until BSIE is less than LoTE or the basis set contribution to calculation time dominates the level of theory contribution. The model chemistry error is related to the basis set incompleteness and level of theory errors by the triangular inequality and is thus at maximum the sum of these two component errors, but can go as low as zero if the basis set incompleteness and level of theory errors cancel one another. 

To calculate these three important properties, we based our analysis in regularised root mean square errors (regularised RMSE) comparing calculated dipole moments ($\mu_{calc} [D]$) with the benchmark values ($\mu_{bmk} [D]$) produced by Hait and Head-Gordon \cite{18HaHe}. These benchmark values were changed accordingly with the property to be determined, i.e. BSIE, LoTE or ModChem Error. The difference between the calculated and benchmark values was regularised as $(\mu_{calc} - \mu_{bmk})/(max(\mu_{bmk},1[D]))$ following the same procedure that Hait and Head-Gordon implemented in their calculations. This choice meant relative errors were calculated for molecules with large dipole molecules but absolute errors were calculated for molecules with small dipole moments.

\subsection{Benchmark values}

Though a true complete basis set calculation is impossible, a near CBS result suitable for comparisons against double and triple zeta basis set results can be easily obtained through very large basis sets or extrapolation from large basis sets. Similarly, CCSD(T) results are a suitable substitute for exact solution to the Schrodinger Equation when considering the accuracy of DFT calculations. 

Our benchmark values were obtained from Hait and Head-Gordon \cite{18HaHe} except for B3LYP-D3. For HF and MP2, they calculated these results from aug-cc-pV\{Q,5\}Z or aug-cc-pV\{T,Q\}Z basis set extrapolation, whereas, for DFT, aug-pc-4 results were used without extrapolation. Following the same procedure, our B3LYP-D3 benchmark values were calculated using aug-pc-4.

\subsection{Computational Details}

All computations were performed using the computational quantum chemistry package Q-Chem 5.2 \cite{15ShGaEp}.

The package Q-Chem 5.2 already includes most of the basis sets tested in this study, except for the ma-$n$ family ($n$ = SVP, TZVP, TZVPP) and the Sadlej pVTZ basis set. These basis sets were defined manually using the general basis command in the input file. The ma-$n$ basis sets were downloaded from the Truhlar research group website at the University of Minnesota (https://comp.chem.umn.edu/basissets/basis.cgi) and the pVTZ basis was downloaded from the Basis Set Exchange website \cite{96Fe,07ScDiEl,19PrAlDi}.

The workflow for the computations was carried out using the ACCDB collection created by Peverati and Morgante \cite{19MoPe}. This collection is based on a Snakemake \cite{12KoRa} flow containing optimised molecular geometries from different databases as well as a set of computational tools useful for file and data handling. The ACCDB collection is supported with a Q-Chem input file template that was modified, when needed, for some calculations. The templates for the input files can be found in SI documentation. 

\section{Results and Discussion}
\label{results}

\subsection{Performance at the CBS Limit}

Our focus here on basis sets necessitates a reduction in the data set size from 152 to 114 molecules after excluding molecules containing Li, Be, B, Na and Al. In this section, we test whether this significantly affects our result. 

\begin{figure}[ht]
    \centering
    \includegraphics[width=0.5\textwidth]{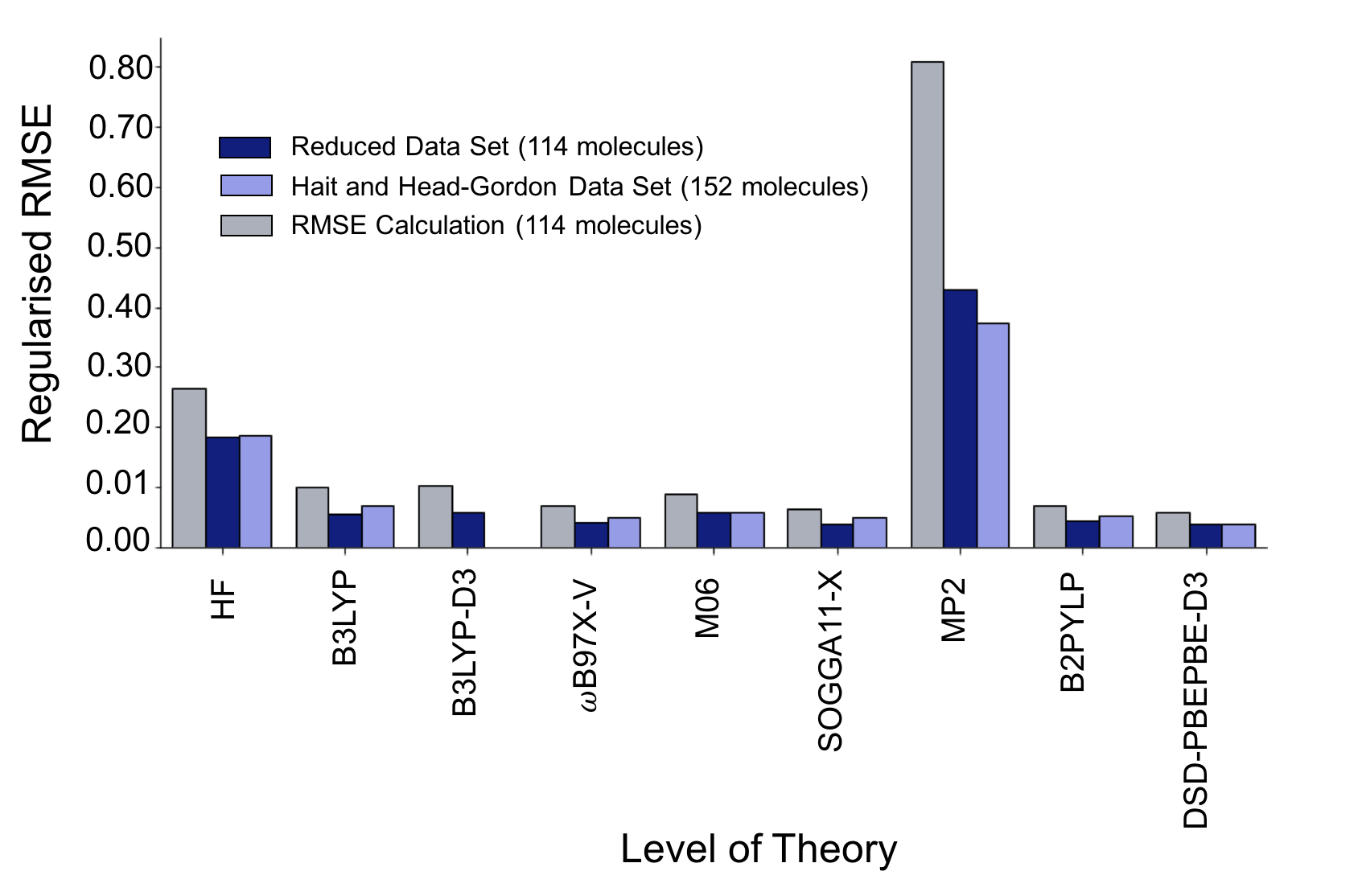}
    \caption{Performance of the chosen levels of theory in the computation of dipole moments. The dark-blue and lilac bar plots correspond to the reduced data set (used in the present study) and the 152 molecules data set, respectively. The grey bar plot corresponds to the reduced data set analysed with regular RMSE instead of regularised RMSE. Apart from B3LYP-D3 (produced here), all data were collected from the data set benchmark study performed by Hait and Head-Gordon \cite{18HaHe}. The $x$ axis in the figure is organised as follows: Hartree-Fock, hybrid functionals, MP2 and double-hybrid functionals.}
    \label{method-error}
\end{figure}

The regularised RMSE values for the reduced data set used in the present study are visually presented in figure \ref{method-error}, along with the regularised RMSE values found by Hait and Head-Gordon using the 152 molecules data set. Figure \ref{method-error} shows that the reduction of the data set does not lead to significant deviations from the overall behaviour delivered by the complete data set. Hybrid and double-hybrid functionals still deliver the best performance, whereas MP2 still accounts for the worst performance due to the broken spin symmetry present in some systems \cite{18HaHe}.

Though dispersion is known to play a crucial role in dipole moment predictions, B3LYP-D3 did not provide any improvement in the computations and was slightly outperformed by the B3LYP functional. 

The grey bar plot in the figure  \ref{method-error} presents the performance of all levels of theory in the reduced data set when no regularisation to the RMSE values is applied. Though the explicit numbers diverge between the data sets and the figures of merit used to perform the analysis (i.e. regularised vs non-regularised RMSE), the overall performance of the different levels of theory remains fairly comparable, indicating that the choice of regularised or non-regularised RMSE values will not have a significant influence in the subsequent analysis. 

\subsection{Finite Basis Set Performance}

Based on the diversity of basis sets assessed in the present study, we decided to classify the basis sets in five different groups according with their augmentation with diffuse and polarisation functions:

\begin{compactenum}
    \item \textit{No Diffuse or Polarisation:} Basis sets with neither diffuse nor polarisation functions, i.e. STO-6G, 6-31G, 6-311G, pc-0 and pcseg-0;
     \item \textit{Diffuse only, no Polarisation:} Basis sets augmented with diffuse functions only, i.e. 6-31+G, 6-311+G , aug-pc-0 and aug-pcseg-0;
    \item \textit{Polarisation, no Diffuse:} Basis sets augmented with polarisation functions only, i.e. 6-31G*, 6-311G*, pc-\{1,2\}, pcseg-\{1,2\}, cc-pV\{D,T\}Z and def2-\{SVP,TZVP,TZVPP\};
   \item \textit{Polarisation \& Diffuse on non-Hydrogen Atoms:} Basis sets augmented with polarisation functions and diffuse function on non-hydrogen atoms only, i.e. 6-31+G*, 6-311+G*, ma-\{SVP,TZVP,TZVPP\}; and
    \item \textit{Polarisation \& Diffuse on Hydrogen and non-Hydrogen Atoms:} Basis sets augmented with polarisation functions and diffuse functions on both hydrogen and non-hydrogen atoms; i.e. 6-31++G**, 6-311++G**, 6-311++G(2df,2pd), aug-pc-\{1,2\}, aug-pcseg-\{1,2\}, aug-cc-pV\{D,T\}Z, def2-\{SVP,TZVP,TZVPP\}D and pVTZ
\end{compactenum}

\begin{figure*}[t!]
    \centering
    \includegraphics[width=0.8\textwidth]{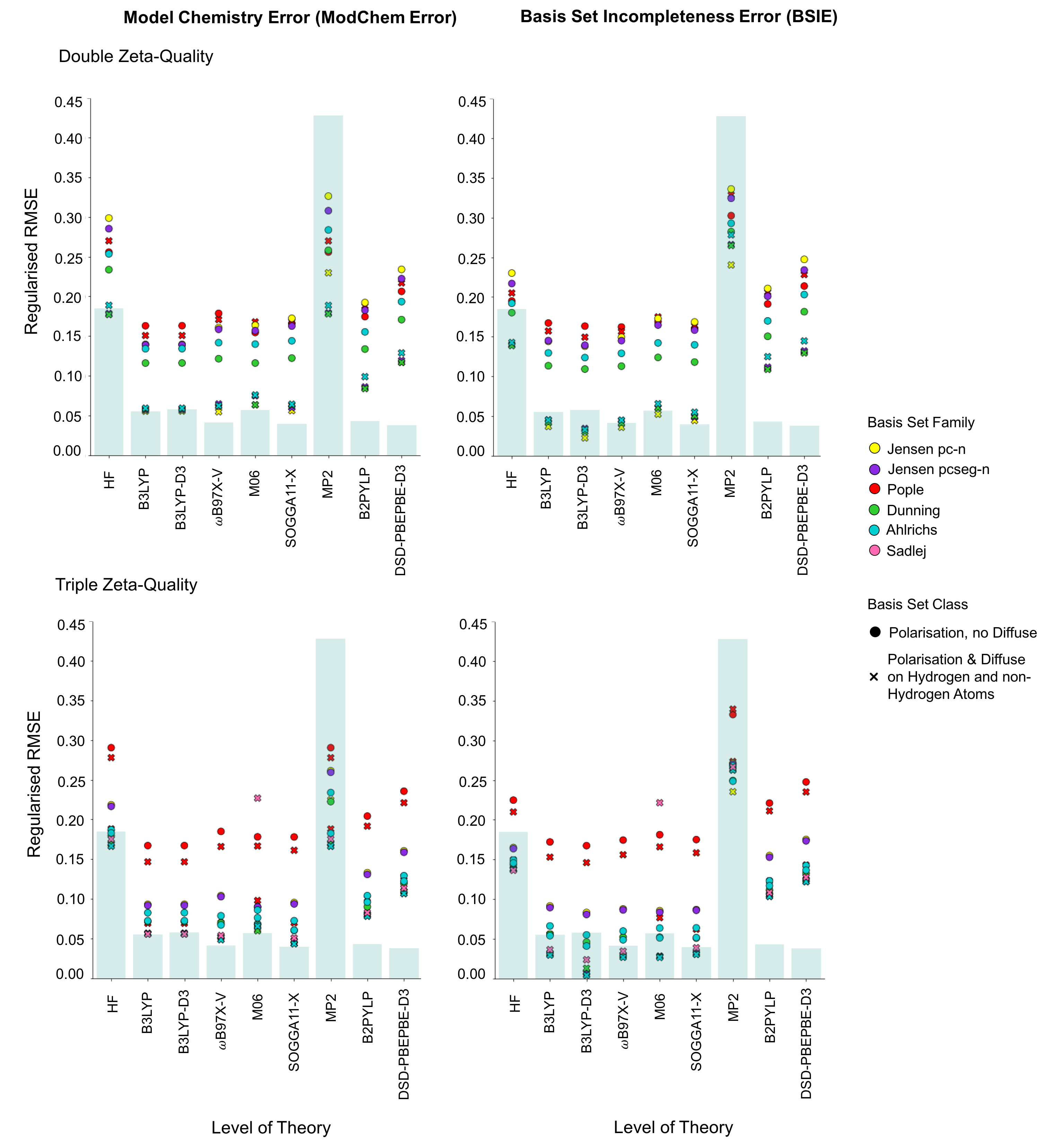}
    \caption{Performance of different basis sets when coupled with the nine levels of theory tested. Each row in the figure corresponds to the zeta-quality of the basis set (double and triple), and the columns correspond to different computational quantum chemistry errors (model chemistry and basis set incompleteness errors, respectively). The bar plots underneath the data points correspond to the level of theory error represented as the dark-blue bar plot in figure \ref{method-error}.}
    \label{all-basis}
\end{figure*}

Tables \ref{tab:rmse-mc} and \ref{tab:rmse-bsi} present the regularised RMSE for the model chemistry (ModChem Error) and basis set incompleteness errors (BSIE) of all basis sets, across the different levels of theory. Figure \ref{all-basis} provides a visual representation of this data including only the \textit{Polarisation, no Diffuse} and \textit{Polarisation \& Diffuse on Hydrogen and non-Hydrogen Atoms} basis set classes. Each row in the figure corresponds to the zeta-quality of the basis set tested. The columns, on the other hand, indicate the different computational quantum chemistry errors: the first column corresponds to the model chemistry error (ModChem Error), whereas the second row corresponds to the basis set incompleteness error (BSIE). The bar plots underneath the data points represent to the level of theory error (LoTE) (dark-blue bar plot presented in figure \ref{method-error}).

\begin{table*}[t!]
\resizebox{\textwidth}{!}{
\centering
     \caption{Regularised RMSE values for all basis sets across the different levels of theory tested in this study. The notation for the augmentation of the basis sets is given at the bottom of the table. The values correspond to the model chemistry error (ModChem Error).}
     \label{tab:rmse-mc}
\begin{tabular}{lccccrrccrcrrcl}
     \toprule
  \textbf{\emph{ModChem Error}} & \mc{4}{c}{\textbf{Augmentation}} & \mc{9}{c}{\textbf{Level of Theory}}  \\
    \cmidrule(r){2-5} \cmidrule(r){6-15}
      \textbf{Basis Set} & \mc{1}{c}{+} & \mc{1}{r}{++} & \mc{1}{c}{*} & \mc{1}{l}{**} & \mc{1}{c}{HF} & \mc{1}{c}{B3LYP} & \mc{1}{c}{B3LYP-D3} & \mc{1}{c}{$\omega$B97X-V} & \mc{1}{c}{M06} & \mc{1}{c}{SOGGA11-X} & \mc{1}{c}{MP2} & \mc{1}{c}{B2PLYP} & \mc{1}{c}{DSD-PBEPBE-D3} & \mc{1}{c}{\textbf{Average}}  \\        
     \midrule
         \vspace{-0.8em} \\
 \textbf{CBS Limit} & & & & & 0.185 & 0.055 & 0.058 & 0.042 & 0.057 & 0.039 & 0.428 & 0.043 & 0.038 & \textbf{0.105} \\
 \textbf{PB3h-3c/def2-mSVP} & & & & &    &   &   &   &   &   &   &   &   & \textbf{0.164} \\
     \vspace{-0.8em} \\
     \vspace{-0.8em} \\
     \mc{2}{l}{Single Zeta-Quality} \\
     \vspace{-0.8em} \\
 \textbf{STO-6G} & \textsf{x} & \textsf{x} & \textsf{x} & \textsf{x} & 0.451 & 0.419 & 0.419 & 0.452 & 0.408 & 0.435 & 0.435 & 0.432 & 0.432 & \textbf{0.431} \\
 \textbf{pc-0} & \textsf{x} & \textsf{x} & \textsf{x} & \textsf{x} & 0.501 & 0.327 & 0.326 & 0.372 & 0.373 & 0.389 & 0.530 & 0.395 & 0.448 & \textbf{0.407} \\
 \textbf{aug-pc-0} & \checkmark & \checkmark & \textsf{x} & \textsf{x} & 0.576 & 0.406 & 0.406 & 0.457 & 0.442 & 0.460 & 0.638 & 0.479 & 0.521 & \textbf{0.487} \\
 \textbf{pcseg-0} & \textsf{x} & \textsf{x} & \textsf{x} & \textsf{x} & 0.491 & 0.328 & 0.328 & 0.384 & 0.360 & 0.384 & 0.551 & 0.383 & 0.439 & \textbf{0.405} \\
 \textbf{aug-pcseg-0} & \checkmark & \checkmark & \textsf{x} & \textsf{x} & 0.551 & 0.391 & 0.391 & 0.445 & 0.412 & 0.438 & 0.511 & 0.455 & 0.499 & \textbf{0.455} \\
     \vspace{-0.8em} \\
    \mc{2}{l}{Double Zeta-Quality} \\
    \vspace{-0.8em} \\
\textbf{6-31G} & \textsf{x} & \textsf{x} & \textsf{x} & \textsf{x} & 0.474 & 0.316 & 0.316 & 0.360 & 0.347 & 0.353 & 0.474 & 0.370 & 0.414 & \textbf{0.380} \\
\textbf{6-31+G} & \checkmark & \textsf{x} & \textsf{x} & \textsf{x} & 0.486 & 0.335 & 0.335 & 0.380 & 0.370 & 0.373 & 0.486 & 0.393 & 0.430 & \textbf{0.399} \\
\textbf{6-31G*} & \textsf{x} & \textsf{x} & \checkmark & \textsf{x} & 0.256 & 0.163 & 0.163 & 0.179 & 0.154 & 0.166 & 0.256 & 0.174 & 0.206 & \textbf{0.191} \\
\textbf{6-31+G*} & \checkmark & \textsf{x} & \checkmark & \textsf{x} & 0.276 & 0.162 & 0.162 & 0.181 & 0.179 & 0.179 & 0.275 & 0.196 & 0.223 & \textbf{0.204} \\
\textbf{6-31++G**} & \checkmark & \checkmark & \checkmark & \checkmark & 0.270 & 0.151 & 0.151 & 0.171 & 0.168 & 0.168 & 0.270 & 0.188 & 0.217 & \textbf{0.195} \\
\textbf{pc-1} & \textsf{x} & \textsf{x} & \checkmark & \checkmark & 0.299 & 0.140 & 0.140 & 0.161 & 0.164 & 0.172 & 0.326 & 0.192 & 0.234 & \textbf{0.203} \\
\textbf{aug-pc-1} & \checkmark & \checkmark & \checkmark & \checkmark & 0.177 & 0.055 & 0.056 & 0.055 & 0.063 & 0.056 & 0.230 & 0.084 & 0.117 & \textbf{0.099} \\
\textbf{pcseg-1} & \textsf{x} & \textsf{x} & \checkmark & \checkmark & 0.285 & 0.139 & 0.139 & 0.159 & 0.157 & 0.163 & 0.308 & 0.182 & 0.222 & \textbf{0.195} \\
\textbf{aug-pcseg-1} & \checkmark & \checkmark & \checkmark & \checkmark & 0.179 & 0.057 & 0.057 & 0.065 & 0.075 & 0.060 & 0.179 & 0.086 & 0.119 & \textbf{0.097} \\
\textbf{cc-pVDZ} & \textsf{x} & \textsf{x} & \checkmark & \checkmark & 0.234 & 0.116 & 0.116 & 0.121 & 0.116 & 0.122 & 0.258 & 0.134 & 0.171 & \textbf{0.154} \\
\textbf{aug-cc-pVDZ} & \checkmark & \checkmark & \checkmark & \checkmark & 0.177 & 0.058 & 0.058 & 0.062 & 0.063 & 0.063 & 0.178 & 0.084 & 0.117 & \textbf{0.096} \\
\textbf{def2-SVP} & \textsf{x} & \textsf{x} & \checkmark & \checkmark & 0.254 & 0.134 & 0.134 & 0.142 & 0.140 & 0.144 & 0.284 & 0.155 & 0.193 & \textbf{0.175} \\
\textbf{ma-SVP} & \checkmark & \textsf{x} & \checkmark & \checkmark & 0.312 & 0.154 & 0.154 & 0.173 & 0.282 & 0.175 & 0.280 & 0.194 & 0.240 & \textbf{0.218} \\
\textbf{def2-SVPD} & \checkmark & \checkmark & \checkmark & \checkmark & 0.189 & 0.059 & 0.059 & 0.062 & 0.076 & 0.064 & 0.189 & 0.099 & 0.129 & \textbf{0.103} \\
    \vspace{-0.8em} \\
    \mc{2}{l}{Triple Zeta-Quality} \\
    \vspace{-0.8em} \\
\textbf{6-311G} & \textsf{x} & \textsf{x} & \textsf{x} & \textsf{x} & 0.472 & 0.316 & 0.316 & 0.360 & 0.340 & 0.346 & 0.472 & 0.375 & 0.414 & \textbf{0.379} \\
\textbf{6-311+G} & \checkmark & \textsf{x} & \textsf{x} & \textsf{x} & 0.469 & 0.323 & 0.323 & 0.368 & 0.350 & 0.354 & 0.469 & 0.380 & 0.415 & \textbf{0.383} \\
\textbf{6-311G*} & \textsf{x} & \textsf{x} & \checkmark & \textsf{x} & 0.290 & 0.167 & 0.167 & 0.185 & 0.178 & 0.178 & 0.290 & 0.204 & 0.236 & \textbf{0.211} \\
\textbf{6-311+G*} & \checkmark & \textsf{x} & \checkmark & \textsf{x} & 0.293 & 0.174 & 0.174 & 0.192 & 0.192 & 0.188 & 0.292 & 0.212 & 0.239 & \textbf{0.217} \\
\textbf{6-311++G**} & \checkmark & \checkmark & \checkmark & \checkmark & 0.278 & 0.147 & 0.147 & 0.166 & 0.167 & 0.161 & 0.278 & 0.192 & 0.221 & \textbf{0.195} \\
\textbf{6-311++G(2df,2pd)} & \checkmark & \checkmark & \checkmark & \checkmark & 0.188 & 0.069 & 0.069 & 0.070 & 0.098 & 0.070 & 0.188 & 0.100 & 0.128 & \textbf{0.109} \\
\textbf{pc-2} & \textsf{x} & \textsf{x} & \checkmark & \checkmark & 0.218 & 0.093 & 0.093 & 0.104 & 0.090 & 0.095 & 0.261 & 0.133 & 0.160 & \textbf{0.139} \\
\textbf{aug-pc-2} & \checkmark & \checkmark & \checkmark & \checkmark & 0.169 & 0.056 & 0.056 & 0.052 & 0.062 & 0.047 & 0.225 & 0.080 & 0.110 & \textbf{0.095} \\
\textbf{pcseg-2} & \textsf{x} & \textsf{x} & \checkmark & \checkmark & 0.216 & 0.092 & 0.092 & 0.103 & 0.090 & 0.094 & 0.259 & 0.131 & 0.159 & \textbf{0.137} \\
\textbf{aug-pcseg-2} & \checkmark & \checkmark & \checkmark & \checkmark & 0.169 & 0.057 & 0.057 & 0.052 & 0.066 & 0.047 & 0.169 & 0.080 & 0.109 & \textbf{0.089} \\
\textbf{cc-pVTZ} & \textsf{x} & \textsf{x} & \checkmark & \checkmark & 0.180 & 0.071 & 0.071 & 0.070 & 0.068 & 0.060 & 0.223 & 0.090 & 0.120 & \textbf{0.106} \\
\textbf{aug-cc-pVTZ} & \checkmark & \checkmark & \checkmark & \checkmark & 0.170 & 0.056 & 0.056 & 0.052 & 0.060 & 0.044 & 0.170 & 0.080 & 0.110 & \textbf{0.088} \\
\textbf{def2-TZVP} & \textsf{x} & \textsf{x} & \checkmark & \checkmark & 0.187 & 0.082 & 0.083 & 0.079 & 0.086 & 0.072 & 0.234 & 0.104 & 0.129 & \textbf{0.117} \\
\textbf{ma-TZVP} & \checkmark & \textsf{x} & \checkmark & \checkmark & 0.234 & 0.076 & 0.076 & 0.097 & 0.234 & 0.073 & 0.234 & 0.129 & 0.176 & \textbf{0.148} \\
\textbf{def2-TZVPD} & \textsf{x} & \textsf{x} & \checkmark & \checkmark & 0.166 & 0.056 & 0.056 & 0.049 & 0.068 & 0.043 & 0.166 & 0.078 & 0.107 & \textbf{0.088} \\
\textbf{def2-TZVPP} & \textsf{x} & \textsf{x} & \checkmark & \checkmark & 0.183 & 0.072 & 0.072 & 0.067 & 0.076 & 0.061 & 0.182 & 0.096 & 0.122 & \textbf{0.104} \\
\textbf{ma-TZVPP} & \checkmark & \textsf{x} & \checkmark & \checkmark & 0.230 & 0.068 & 0.068 & 0.088 & 0.230 & 0.061 & 0.182 & 0.122 & 0.171 & \textbf{0.136} \\
\textbf{def2-TZVPPD} & \checkmark & \checkmark & \checkmark & \checkmark & 0.166 & 0.056 & 0.056 & 0.048 & 0.066 & 0.043 & 0.166 & 0.078 & 0.107 & \textbf{0.088} \\
\textbf{pVTZ} & \checkmark & \checkmark & \checkmark & \checkmark & 0.175 & 0.056 & 0.056 & 0.054 & 0.227 & 0.051 & 0.175 & 0.083 & 0.113 & \textbf{0.110} \\
    \vspace{-0.8em} \\
     \bottomrule
     \end{tabular}
\begin{tabular}{ccl}
\\
             Column        &     Notation                  &       \\
\midrule
   2  & + &   Diffuse functions on non-hydrogen atoms only  \\
   3  & ++ &  Diffuse functions on both hydrogen and non-hydrogen atoms \\ 
   4  & * &   Polarisation functions on non-hydrogen atoms only   \\
   5  & ** &  Polarisation functions on both hydrogen and non-hydrogen atoms \\
\bottomrule
\end{tabular}
}
\end{table*}

\begin{table*}[t!]
\resizebox{\textwidth}{!}{
\centering
     \caption{Regularised RMSE values for all basis sets across the different levels of theory tested in this study. The notation for the augmentation of the basis sets is given at the bottom of the table. The values correspond to the basis set incompleteness error (BSI Error).}
     \label{tab:rmse-bsi}
\begin{tabular}{lccccrrccrcrrcl}
     \toprule
  \textbf{\emph{BSIE}} & \mc{4}{c}{\textbf{Augmentation}} & \mc{9}{c}{\textbf{Level of Theory}}  \\
    \cmidrule(r){2-5} \cmidrule(r){6-15}
      \textbf{Basis Set} & \mc{1}{c}{+} & \mc{1}{r}{++} & \mc{1}{c}{*} & \mc{1}{l}{**} & \mc{1}{c}{HF} & \mc{1}{c}{B3LYP} & \mc{1}{c}{B3LYP-D3} & \mc{1}{c}{$\omega$B97X-V} & \mc{1}{c}{M06} & \mc{1}{c}{SOGGA11-X} & \mc{1}{c}{MP2} & \mc{1}{c}{B2PLYP} & \mc{1}{c}{DSD-PBEPBE-D3} & \mc{1}{c}{\textbf{Average}}  \\        
     \midrule
         \vspace{-0.8em} \\
\textbf{CBS Limit} & & & & & 0.185 & 0.055 & 0.058 & 0.042 & 0.057 & 0.039 & 0.428 & 0.043 & 0.038 & \textbf{0.105} \\
    \vspace{-0.8em} \\
    \mc{2}{l}{Single Zeta-Quality} \\
    \vspace{-0.8em} \\
\textbf{STO-6G} & \textsf{x} & \textsf{x} & \textsf{x} & \textsf{x} & 0.459 & 0.421 & 0.419 & 0.453 & 0.413 & 0.438 & 0.473 & 0.435 & 0.433 & \textbf{0.438} \\
\textbf{pc-0} & \textsf{x} & \textsf{x} & \textsf{x} & \textsf{x} & 0.422 & 0.340 & 0.336 & 0.360 & 0.393 & 0.386 & 0.524 & 0.412 & 0.458 & \textbf{0.403} \\
\textbf{aug-pc-0} & \checkmark & \checkmark & \textsf{x} & \textsf{x} & 0.479 & 0.423 & 0.418 & 0.446 & 0.465 & 0.458 & 0.608 & 0.496 & 0.532 & \textbf{0.481} \\
\textbf{pcseg-0} & \textsf{x} & \textsf{x} & \textsf{x} & \textsf{x} & 0.407 & 0.341 & 0.337 & 0.368 & 0.386 & 0.378 & 0.580 & 0.397 & 0.443 & \textbf{0.404} \\
\textbf{aug-pcseg-0} & \checkmark & \checkmark & \textsf{x} & \textsf{x} & 0.457 & 0.409 & 0.403 & 0.432 & 0.438 & 0.434 & 0.513 & 0.472 & 0.507 & \textbf{0.452} \\
    \vspace{-0.8em} \\
    \mc{2}{l}{Double Zeta-Quality} \\
    \vspace{-0.8em} \\
\textbf{6-31G} & \textsf{x} & \textsf{x} & \textsf{x} & \textsf{x} & 0.387 & 0.331 & 0.328 & 0.348 & 0.372 & 0.350 & 0.498 & 0.387 & 0.422 & \textbf{0.380} \\
\textbf{6-31+G} & \checkmark & \textsf{x} & \textsf{x} & \textsf{x} & 0.397 & 0.351 & 0.346 & 0.370 & 0.392 & 0.371 & 0.521 & 0.409 & 0.439 & \textbf{0.400} \\
\textbf{6-31G*} & \textsf{x} & \textsf{x} & \checkmark & \textsf{x} & 0.195 & 0.167 & 0.163 & 0.162 & 0.169 & 0.160 & 0.302 & 0.191 & 0.214 & \textbf{0.192} \\
\textbf{6-31+G*} & \checkmark & \textsf{x} & \checkmark & \textsf{x} & 0.210 & 0.169 & 0.161 & 0.166 & 0.187 & 0.175 & 0.331 & 0.213 & 0.234 & \textbf{0.205} \\
\textbf{6-31++G**} & \checkmark & \checkmark & \checkmark & \checkmark & 0.205 & 0.157 & 0.149 & 0.156 & 0.175 & 0.164 & 0.329 & 0.206 & 0.228 & \textbf{0.197} \\
\textbf{pc-1} & \textsf{x} & \textsf{x} & \checkmark & \checkmark & 0.230 & 0.144 & 0.138 & 0.149 & 0.173 & 0.168 & 0.336 & 0.211 & 0.247 & \textbf{0.200} \\
\textbf{aug-pc-1} & \checkmark & \checkmark & \checkmark & \checkmark & 0.140 & 0.037 & 0.022 & 0.036 & 0.052 & 0.044 & 0.240 & 0.110 & 0.131 & \textbf{0.090} \\
\textbf{pcseg-1} & \textsf{x} & \textsf{x} & \checkmark & \checkmark & 0.217 & 0.145 & 0.139 & 0.145 & 0.165 & 0.158 & 0.324 & 0.201 & 0.234 & \textbf{0.192} \\
\textbf{aug-pcseg-1} & \checkmark & \checkmark & \checkmark & \checkmark & 0.139 & 0.045 & 0.034 & 0.044 & 0.060 & 0.049 & 0.266 & 0.112 & 0.132 & \textbf{0.098} \\
\textbf{cc-pVDZ} & \textsf{x} & \textsf{x} & \checkmark & \checkmark & 0.180 & 0.113 & 0.109 & 0.113 & 0.124 & 0.118 & 0.282 & 0.150 & 0.181 & \textbf{0.152} \\
\textbf{aug-cc-pVDZ} & \checkmark & \checkmark & \checkmark & \checkmark & 0.139 & 0.042 & 0.030 & 0.041 & 0.060 & 0.050 & 0.265 & 0.109 & 0.129 & \textbf{0.096} \\
\textbf{def2-SVP} & \textsf{x} & \textsf{x} & \checkmark & \checkmark & 0.192 & 0.130 & 0.124 & 0.129 & 0.142 & 0.140 & 0.293 & 0.170 & 0.203 & \textbf{0.169} \\
\textbf{ma-SVP} & \checkmark & \textsf{x} & \checkmark & \checkmark & 0.244 & 0.160 & 0.152 & 0.160 & 0.285 & 0.171 & 0.340 & 0.213 & 0.250 & \textbf{0.219} \\
\textbf{def2-SVPD} & \checkmark & \checkmark & \checkmark & \checkmark & 0.143 & 0.045 & 0.033 & 0.045 & 0.065 & 0.055 & 0.278 & 0.125 & 0.145 & \textbf{0.104} \\
    \vspace{-0.8em} \\
    \mc{2}{l}{Triple Zeta-Quality} \\
    \vspace{-0.8em} \\
\textbf{6-311G} & \textsf{x} & \textsf{x} & \textsf{x} & \textsf{x} & 0.386 & 0.332 & 0.328 & 0.351 & 0.361 & 0.345 & 0.498 & 0.393 & 0.424 & \textbf{0.380} \\
\textbf{6-311+G} & \checkmark & \textsf{x} & \textsf{x} & \textsf{x} & 0.382 & 0.339 & 0.333 & 0.358 & 0.370 & 0.352 & 0.507 & 0.397 & 0.424 & \textbf{0.385} \\
\textbf{6-311G*} & \textsf{x} & \textsf{x} & \checkmark & \textsf{x} & 0.225 & 0.172 & 0.167 & 0.174 & 0.181 & 0.175 & 0.333 & 0.221 & 0.248 & \textbf{0.211} \\
\textbf{6-311+G*} & \checkmark & \textsf{x} & \checkmark & \textsf{x} & 0.224 & 0.179 & 0.172 & 0.180 & 0.193 & 0.184 & 0.349 & 0.229 & 0.252 & \textbf{0.218} \\
\textbf{6-311++G**} & \checkmark & \checkmark & \checkmark & \checkmark & 0.210 & 0.153 & 0.146 & 0.156 & 0.166 & 0.158 & 0.339 & 0.211 & 0.235 & \textbf{0.197} \\
\textbf{6-311++G(2df,2pd)} & \checkmark & \checkmark & \checkmark & \checkmark & 0.149 & 0.056 & 0.042 & 0.051 & 0.077 & 0.062 & 0.273 & 0.123 & 0.143 & \textbf{0.108} \\
\textbf{pc-2} & \textsf{x} & \textsf{x} & \checkmark & \checkmark & 0.165 & 0.091 & 0.083 & 0.088 & 0.085 & 0.087 & 0.272 & 0.155 & 0.175 & \textbf{0.133} \\
\textbf{aug-pc-2} & \checkmark & \checkmark & \checkmark & \checkmark & 0.138 & 0.031 & 0.007 & 0.029 & 0.028 & 0.033 & 0.235 & 0.105 & 0.125 & \textbf{0.081} \\
\textbf{pcseg-2} & \textsf{x} & \textsf{x} & \checkmark & \checkmark & 0.164 & 0.089 & 0.081 & 0.086 & 0.083 & 0.086 & 0.270 & 0.153 & 0.173 & \textbf{0.132} \\
\textbf{aug-pcseg-2} & \checkmark & \checkmark & \checkmark & \checkmark & 0.138 & 0.031 & 0.007 & 0.029 & 0.028 & 0.032 & 0.265 & 0.105 & 0.124 & \textbf{0.084} \\
\textbf{cc-pVTZ} & \textsf{x} & \textsf{x} & \checkmark & \checkmark & 0.143 & 0.056 & 0.046 & 0.052 & 0.052 & 0.051 & 0.249 & 0.112 & 0.133 & \textbf{0.099} \\
\textbf{aug-cc-pVTZ} & \checkmark & \checkmark & \checkmark & \checkmark & 0.137 & 0.032 & 0.013 & 0.030 & 0.028 & 0.031 & 0.264 & 0.105 & 0.124 & \textbf{0.085} \\
\textbf{def2-TZVP} & \textsf{x} & \textsf{x} & \checkmark & \checkmark & 0.149 & 0.066 & 0.055 & 0.060 & 0.064 & 0.064 & 0.249 & 0.123 & 0.142 & \textbf{0.108} \\
\textbf{ma-TZVP} & \checkmark & \textsf{x} & \checkmark & \checkmark & 0.192 & 0.072 & 0.060 & 0.079 & 0.225 & 0.062 & 0.293 & 0.145 & 0.185 & \textbf{0.146} \\
\textbf{def2-TZVPD} & \textsf{x} & \textsf{x} & \checkmark & \checkmark & 0.138 & 0.030 & 0.004 & 0.027 & 0.027 & 0.030 & 0.262 & 0.103 & 0.122 & \textbf{0.083} \\
\textbf{def2-TZVPP} & \textsf{x} & \textsf{x} & \checkmark & \checkmark & 0.146 & 0.054 & 0.041 & 0.049 & 0.051 & 0.052 & 0.268 & 0.117 & 0.137 & \textbf{0.102} \\
\textbf{ma-TZVPP} & \checkmark & \textsf{x} & \checkmark & \checkmark & 0.190 & 0.054 & 0.038 & 0.073 & 0.221 & 0.050 & 0.270 & 0.140 & 0.181 & \textbf{0.135} \\
\textbf{def2-TZVPPD} & \checkmark & \checkmark & \checkmark & \checkmark & 0.137 & 0.029 & 0.004 & 0.027 & 0.026 & 0.030 & 0.263 & 0.103 & 0.122 & \textbf{0.082} \\
\textbf{pVTZ} & \checkmark & \checkmark & \checkmark & \checkmark & 0.136 & 0.036 & 0.024 & 0.034 & 0.221 & 0.039 & 0.267 & 0.108 & 0.128 & \textbf{0.110} \\
    \vspace{-0.8em} \\
     \bottomrule
     \end{tabular}
\begin{tabular}{ccl}
\\
             Column        &     Notation                  &       \\
\midrule
   2  & + &   Diffuse functions on non-hydrogen atoms only  \\
   3  & ++ &  Diffuse functions on both hydrogen and non-hydrogen atoms \\ 
   4  & * &   Polarisation functions on non-hydrogen atoms only   \\
   5  & ** &  Polarisation functions on both hydrogen and non-hydrogen atoms \\
\bottomrule
\end{tabular}
}
\end{table*}

It is clear that for most double and triple zeta basis sets, the basis set incompleteness error is larger than the level of theory error, demonstrating that for practical calculations of dipole moment, the basis set choice is at least as important as choice of density functional approximation. In most cases, the model chemistry error and basis set incompleteness error are very similar; only for some hybrid functionals with triple zeta basis sets does the level of theory error dominate over the basis set incompleteness error in determining the overall model chemistry error. There is no  significant evidence of large scale error cancellation between basis set incompleteness error and level of theory error. Even for the oft-utilised B3LYP/6-31G*, there is very minimal error cancellation and the basis set incompleteness error dominates: model chemistry error is 0.163, basis set incompleteness error is 0.167 and the level of theory (i.e. DFT) error is 0.055.

Figure \ref{all-basis} also shows that the performance of a given basis set is significantly enhanced by augmentation with diffuse functions on both hydrogen and non-hydrogen atoms. These bases led to the lowest regularised RMSE values when compared with their versions without diffuse functions, or with diffuse functions on non-hydrogen atoms only (see tables \ref{tab:rmse-mc} and \ref{tab:rmse-bsi}). In fact, when diffuse functions are only included on non-hydrogen atoms, the performance of the basis set is considerably worse. This is attributed to the fact that hydrogen atoms are often placed in the outer region of molecules, which must be accurately described to predict reliable dipole moments.

Overall, the double hybrid functionals perform noticeably poorer with finite basis sets than the hybrid functionals, despite the MP2 finite basis set results being superior to the MP2 infinite basis set results. The reasons for this result are unclear to the authors, but notable in terms of directing practical recommendation for calculation types especially as double hybrids are much more computationally expensive than hybrid functionals.

Though our results show that the performance of double and triple-zeta basis sets are overall clearly superior to single-zeta basis sets, the choice of exactly which double or triple-zeta basis set is used also substantially affects the quality of a dipole moment calculation. In fact, based on our results, the choice of the basis set is far more important than the choice of DFT functional for practical dipole moment calculations where very large basis sets are unsuitable.

We therefore consider in turn the best basis set choice of single, double and triple-zeta quality; the zeta quality is of course a good proxy for calculation time. 

\textbf{Single-zeta basis sets:} Tables \ref{tab:rmse-mc} and \ref{tab:rmse-bsi} clearly show that the single-zeta basis sets introduce basis set incompleteness and model chemistry errors far in excess of the level of theory error and thus should be avoid in dipole moment computations. Augmentation with diffuse functions on both hydrogen and non-hydrogen atoms does not improve the performance, i.e. aug-pc-0 and aug-pcseg-0 do not outperform other single-zeta basis sets. This demonstrates the crucial role that the basis set zeta-quality plays in electronic structure calculations. Although the computation of dipole moments depends upon an accurate description of the tail region of the wave function, and therefore augmenting the basis set with diffuse functions seems reasonable, the basis set used needs to provide an accurate electronic representation first, in order to deliver reliable results.

\textbf{Double Zeta-Quality Basis Sets:} 
The Jensen pc-1 and pcseg-1, the Ahlrichs-Karlsruhe def2-SVP and Dunning cc-pVDZ bases have very good performance, with regularised RMSE between 0.152 and 0.203 on average. The augmented Jensen and Dunning double-zeta basis sets, i.e. aug-pc-1 and aug-pcseg-1 and aug-cc-pVDZ, perform even better, with BSIE and ModChem errors approaching and even sometimes dropping below the infinite basis set level of theory error (LoTE) for hybrid functionals. The def2-SVPD basis has an slightly superior performance than the unaugmented def2-SVP, whereas ma-SVP -- augmented only on non-hydrogen atoms -- performs considerably worse.
Notably, the Pople-style basis sets of double zeta-quality account for the worst performance of all double zeta-quality basis sets tested. Neither the augmentation with diffuse (6-31+G) nor polarisation (6-31G*) functions on non-hydrogen atoms only results in any improvement in the results.  The 6-31+G* and 6-31++G** bases set deliver a significant improvement, but are still outperformed by the Jensen and Dunning basis sets. 

The low-cost electronic structure approach PBEh-3c/def2-mSVP \cite{15GrBrBa} is known to provide fast accurate general chemical properties by means of error cancellation between BSI and level of theory errors. The regularised RMSE for this approach, when used for the calculation of dipole moments, is 0.164 (second row in table \ref{tab:rmse-mc}) which makes it comparable with the average performance of the Jensen, Dunning and Ahlrichs-Karlsruhe unaugmented double-zeta basis sets. The model also outperforms most of the widely used B3LYP/6-31G and B3LYP/6-31+G models and its performance is comparable to the B3LYP/6-31G* model. However, when double-zeta bases augmented with diffuse functions on both hydrogen and non-hydrogen atoms are used, the PBEh-3c/def2-mSVP model is significantly outperformed.

\textbf{Triple Zeta-Quality Basis Sets}: Once again, the Jensen, Dunning and Ahlrichs-Karlsruhe basis sets account for the best overall performance with augmented functions reducing basis set incompleteness errors by about 20 to 40\% relative to unaugmented functions. The Sadlej pVTZ basis set also has very strong performance, with the performance of the unaugmented Ahlrichs-Karlsruhe basis sets (def2-TZVP and def2-TZVPP) being similar. Finally, most of the Pople-style 6-311G bases, and its further augmentations, have the worst overall performance and do not improve overall to the double-zeta Pople 6-31G family. This poor performance can be attributed to the fact that Pople's 6-311G family basis functions are too tight and the extra basis set flexibility acts to give flexibility to the description of the core and not valence electrons \cite{20CoZaMc}. The 6-311++G(2df,2pd) basis, with additional diffuse and polarisation functions on both hydrogen and non-hydrogen atoms, provides a significant improvement in the calculations, with regularised RMSE values similar to the unaugmented Jensen, Dunning and Ahlrichs-Karlsruhe bases, but it is still outperformed by the augmented version of the aforementioned basis sets.

\subsection{Spin-Polarised and Non-Spin-Polarised Molecules:}

\begin{figure}[ht]
    \centering
    \includegraphics[width=0.5\textwidth]{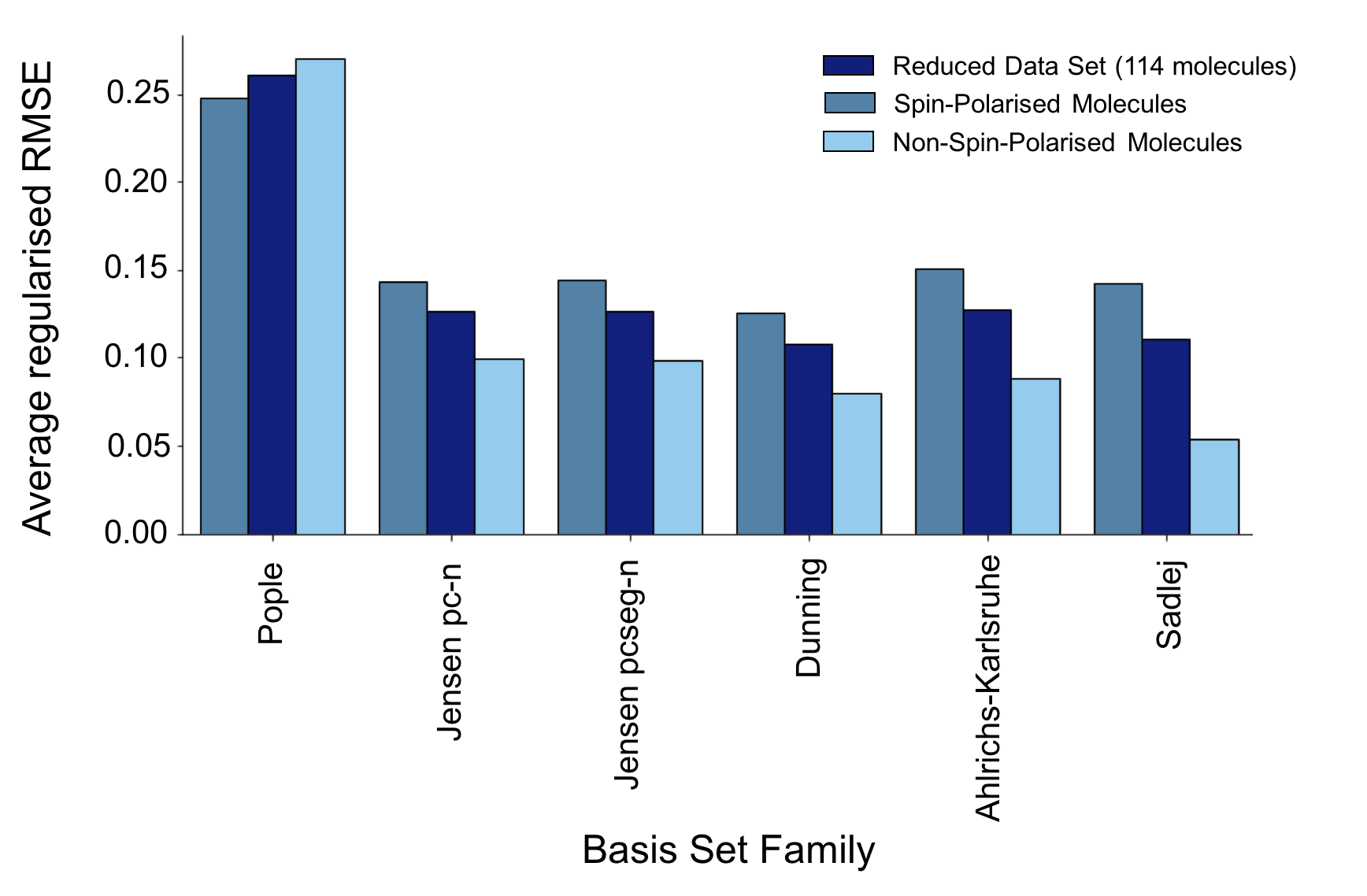}
    \caption{Average performance of the different basis set families when considering only spin-polarised and non-spin-polarised molecules. The average performance across the reduced data set is also presented.}
    \label{sp-nsp}
\end{figure}

Open-shell (spin-polarised) molecules often represent a bigger computational challenge than closed-shell molecules (non-spin-polarised) \cite{99BaBo}. This requirement leads, in some cases, to different recommendations regarding level of theory and basis set choice. Here, we have divided the data set into spin-polarised and non-spin-polarised molecules to see the influence of the basis set choice in these two groups. Figure \ref{sp-nsp} presents the average performance of the different basis set families when considering spin-polarised and non-spin-polarised molecules only. The average performance was taken over double and triple-zeta quality bases only, as our results show that they provide the strongest performance in the computations.

Apart from the Pople family, the figure shows that a better performance for a given basis set will be found when the computations involve non-spin-polarised molecules. However, the trend across the different groups of molecules remains fairly unchanged, indicating that there is no preference of basis set choice when considering spin-polarised and non-spin-polarised molecules. This implies that the recommendations given from the 114 molecules analysis can also be generalised to these two groups of molecules.

\subsection{Recommendations for Users}

Our results clearly show that Pople-style basis sets, e.g. 6-31G*, are substantially outperformed by Jensen, Dunning and Ahlrichs-Karlsruhe basis sets when calculating dipole moments and should not be considered among the options for these calculations. The widely used B3LYP/6-31G* model could be replaced with the PBEh-3c/def2-mSVP model as a much faster method with similar performance.

For a very high accuracy calculation for small molecules, double hybrid functional calculations should be paired with a quadrupole basis set or CBS-extrapolation, following the recommendations of Hait and Head-Gordon \cite{18HaHe}. 

For larger molecules or for studies with a large number of target molecules, we recommend a hybrid functional paired with the aug-pcseg-1 or aug-pc-1 basis sets. From the results of Hait and Head-Gordon \cite{18HaHe}, $\omega$B97X-V and SOGGA11-X are good choices since larger basis set calculations (e.g. aug-pcseg-2) can reliably be used to benchmark selected values if desired. The Jensen basis sets are recommended over the Dunning basis sets as the former were specifically optimised for DFT calculations and come in segmented and unsegmented versions to be suitable for different integral evaluation packages. 

\begin{figure*}[t!]
    \centering
    \includegraphics[width=1\textwidth]{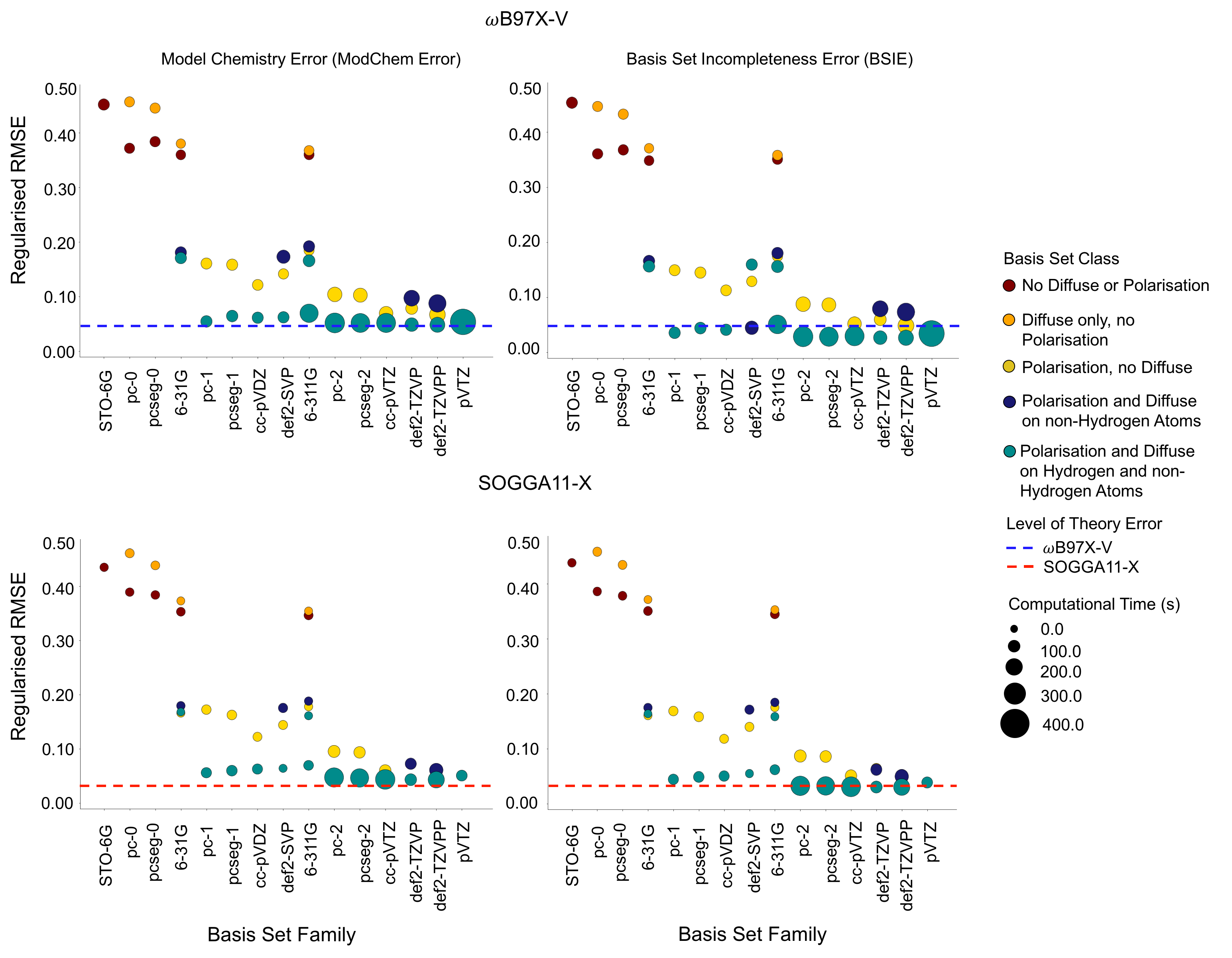}
    \caption{Performance for all basis sets at the $\omega$B97X-V and SOGGA11-X levels of theory (first and second row in the figure, respectively). The dotted line in each plot represents the respective level of theory error. The first column corresponds to the model chemistry error (ModChem Error) and the second column to the basis set incompleteness error (BSIE). The size of the data points in the figure represents the computational time required to perform the calculations with the most time-consuming molecule (CH$_3$O) from the data set with Q-Chem 5.2. Note that calculation times are very sensitive to program and platform, and are timings that should only be considered as indicative.}
    \label{hyb-rmse}
\end{figure*}

We justify our recommendations with figure \ref{hyb-rmse}, which compares the performance of all basis sets tested when coupled with the best two hybrid functionals, $\omega$B97X-V and SOGGA11-X, in terms of approximate timings and errors. It is clear and aligns with our expectation that the triple-zeta basis sets (augmented and unaugmented) are much more time-consuming than the augmented double-zeta basis sets. Yet, compared to the augmented double zeta basis sets, the unaugmented triple zeta basis sets perform worse and the augmented triple zeta basis sets only slightly better, justifying our earlier recommendations. 

It is worth noting that the Sadlej pVTZ basis, specifically optimised for dipole moments, can provide an outstanding performance in the calculation of dipole moments. However, one must conduct a careful selection of the hybrid functional chosen to perform the calculations as, in some cases, the coupling could result in a more time-consuming calculation (e.g. $\omega$B97X-V/pVTZ in figure \ref{hyb-rmse}) or derive in less accurate results (e.g. M06/pVTZ in tables \ref{tab:rmse-mc} and \ref{tab:rmse-bsi}). This erratic performance can perhaps be attributed to the fact the original basis set design in 1988-1992 was for high level correlated methods as DFT functionals were not yet widely available. Furthermore, the Sadlej basis set is not available for some elements. Therefore, we are hesitant to recommend this basis set currently, though we support the overall practice of designing basis sets for specific applications. 

\section{Concluding Remarks}
\label{conclusion}

To provide an effective yet reliable way to computationally predict dipole moments, here we have tested the performance of 38 general-purpose basis sets, of single up to triple zeta quality, in the computation of dipole moments. The basis sets were coupled with nine different levels of theory, belonging to hybrid and double-hybrid density functionals, and wave function-based methods.

Our results show that the best compromise between accuracy and computational efficiency is achieved by coupling a hybrid functional (e.g. $\omega$B97X-V or SOGGA11-X) with an augmented double zeta-quality basis from the Jensen (aug-pc-1 or aug-pcseg-1) or Dunning (aug-cc-pVDZ) families. One could obtain a slightly superior performance by using an augmented triple zeta-quality basis set, but the computational cost makes it impractical. 

Hait and Head-Gordon \cite{18HaHe} suggested that double-hybrid functionals provide the best way to predict dipole moments. However, our results show that it is only the case when the computations involve CBS limit results; double hybrid functionals were outperformed by hybrid functionals when they both were coupled with double and triple zeta-quality basis sets.

The triple-zeta Sadlej pVTZ basis can also provide reliable and fast results. However, probably because it was designed for correlated methods, we found it has some erratic performance with certain hybrid functionals, e.g. lower performance or significantly larger calculation times. Nevertheless, basis sets designed for specific properties have significant promise in enabling high performance for reduced cost. 

Single zeta-quality and Pople-style basis sets should not be implemented in calculations involving dipole moments. These basis sets presented the worst regularised RMSE values across all different levels of theory.

Finally, our results highlight the relevant role played by diffuse functions in the computation of dipole moments. The augmentation of the basis sets with diffuse functions on both hydrogen and non-hydrogen atoms results in an overall enhanced performance when compared with the unaugmented basis sets.

\begin{acknowledgement}

This research was undertaken with the assistance of resources from the National Computational Infrastructure (NCI Australia), an NCRIS enabled capability supported by the Australian Government.

The authors declare no conflicts of interest. 

\end{acknowledgement}

\begin{suppinfo}

The data underlying this paper are available in the paper itself and in its online supplementary material. These include the following files and folders;
\begin{itemize}
\itemsep0em
    \item \texttt{si\_documentation.pdf} - explanation of the files and workflow within the SI documentation. This PDF also contains the input file sample used to perform the calculations.
    \item \texttt{dipole\_analysis.xlsx} - data analysis for the different method/basis set dipole moment calculations.
    \item \texttt{comp\_times.xlsx} - timings for each method/basis set combination.
    \item \texttt{NSP-dataset.xlsx} - data analysis for non-spin-polarised molecules only.
    \item \texttt{SP-data.xlsx} - data analysis for spin-polarised molecules only.
    \item \textsf{Basis\_LoT\_Performace} - folder containing csv files with the relevant information from each output file.
    \item \textsf{geometries\_dipolemoment} - folder containing all reference geometries used throughout the study.
\end{itemize}

\end{suppinfo}

\bibliography{references}

\end{document}